\documentclass[prc,aps,showpacs,nofootinbib,twocolumn]{revtex4}
\pdfoutput=1 
\usepackage{amsmath}
\usepackage{amsfonts}
\usepackage{graphicx}
\usepackage{dcolumn}
\usepackage{hyperref}
\newcommand{\fpf}[2]{{F}_{#1}^{*}{F}_{#2}}

\begin{document}

\include{defs}

\title{Model dependence of single-energy fits to pion photoproduction data}
\author{Ron L.\ Workman, Mark W.\ Paris, William J.\ Briscoe,\,$^a$ \\
Lothar Tiator, Sven Schumann, Michael Ostrick$\,^b$,\\
and Sabit S.\ Kamalov$\,^c$}
\affiliation{ $^a$Center for Nuclear Studies,
Department of Physics, The George Washington University,
Washington, D.C. 20052 \\
$^b$Institut f\"ur Kernphysik,
Johannes Gutenberg Universit\"at, D-55099, Mainz, Germany \\
$^c$Bogoliubov Laboratory for Theoretical Physics, JINR Dubna,
141980 Moscow Region, Russia}

\date{\today}

\begin{abstract}
Model dependence of multipole analysis has been explored through
energy-dependent and single-energy fits to pion photoproduction data. The MAID
energy-dependent solution has been used as input for an event generator
producing realistic pseudo data. These were fitted using the SAID
parametrization approach to determine single-energy and energy-dependent
solutions over a range of lab photon energies from 200 to 1200 MeV. The resulting
solutions were found to be consistent with the input amplitudes from MAID.
Fits with a $\chi$-squared per
datum of unity or less were generally achieved. We discuss energy regions where
consistent results are expected, and explore the sensitivity of fits to the
number of included single- and double-polarization observables. The influence
of Watson's theorem is examined in detail.
\end{abstract}


\maketitle

\section{Introduction}
\label{sec:intro} The non-perturbative regime of quantum chromodynamics (QCD)
is characterized by an array of hadronic resonances. The gross features of this
spectrum have been understood theoretically within the context of constituent
quark models. Experimentally, measurements of reaction observables from an
array of collision processes, in particular pion photoproduction, have shown
the limitations of the quark model description. Precision electromagnetic
facilities around the world have begun to give a more detailed picture of the
hadronic resonances, particularly for the resonances of the nucleon, and
ushered in a renaissance in the field of hadronic reaction theory.

Crucial to this program is the determination of reaction amplitudes from
experimental observables. Model independent extraction of reaction amplitudes
in both spin (\textit{eg.} CGLN, helicity, transversity, \ldots) and
partial-wave bases is a well-studied yet complex task. Typically,
coupled-channel models based on Lagrangians of hadronic effective field theory
involve hundreds of bare parameters in order to obtain realistic descriptions
of data covering the first, second, and third resonance regions. Complementary
to this approach is the extraction of amplitudes with smaller numbers of
parameters and minimal model dependence.

Two types of analysis, typically performed in parametrizing the reaction
amplitudes, lead to energy dependent (ED) or global solutions and energy
independent or single-energy (SE) solutions. The SE amplitudes, generally
extracted through the partial-wave analysis of scattering or reaction data, are
often used as the starting point for more involved multi-channel
analyses~\cite{ses_data}. The term {\it data} is commonly applied to these sets
of amplitudes, implying a relatively model-independent link to the underlying
experimental data. This approach has been successful for some hadronic and
nuclear collision processes. For example, complete or nearly complete sets of
data have been assembled over restricted kinematic regions for
nucleon-nucleon~\cite{nn_ses} and pion-nucleon~\cite{pn_ses} elastic
scattering.  These have allowed a formal study of the ambiguities associated
with the amplitude reconstruction process and yielded a set of useful
amplitudes\cite{Arndt:2006bf} employed, for example, in the determination of
the isospin $\tfrac{1}{2}$ $N$ and isospin-$\tfrac{3}{2}$ $\Delta$ nucleon
resonances\cite{RPP}.

\begin{figure*}
\includegraphics[width=0.60\textwidth]{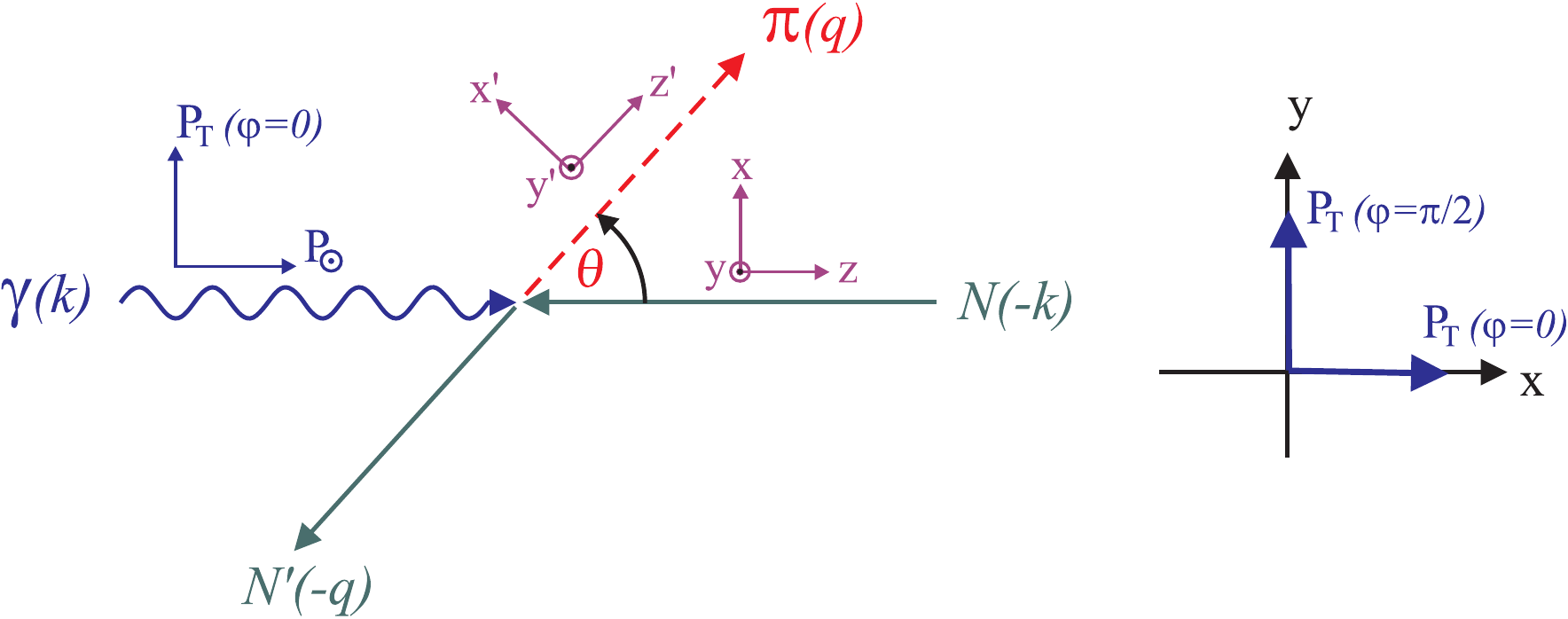}
\caption{\label{fig:frames} Frames for polarization vectors in the CM.}
\end{figure*}

In pion photoproduction, which is the focus of the present study, SE solutions
are generally determined from data within a narrow range of energies, with less
than the number of observables required for a direct, unique amplitude
reconstruction. In order to obtain a stable solution, further constraints are
required. Generally, the SE solutions or fits
are constrained by the results of a global ED fit. Below, we will discuss how
the ED and SE fits are related, and how this procedure differs from amplitude
reconstruction.

Multipole fits from the MAID and SAID groups have been updated to include
improved measurements as they have become available.
Both ED and SE results have been presented and some multipoles show significant
differences. In order to avoid the influence of systematic uncertainties
in the database, as a
potential source of these discrepancies, we have generated pseudo data based on
the MAID and SAID ED fits. Multipoles extracted from these idealized datasets
show much less variation than is seen in fits to the existing experimental
datasets. Below we compare multipoles analyzed using the two different
techniques, ED and SE, and suggest energy ranges over which varying degrees of
model independence can be expected.

In the next section, we provide information sufficient to fix the
phase and sign conventions of amplitudes and observables.
It is important to note that the phase and sign conventions
adopted by the MAID and SAID groups have been consistent for many
years and, as such, have been adopted by a large number of
experimental groups. Recent
publications\cite{Sandorfi:2010uv,Dey:2010fb}, however, have departed
from this convention and we offer the complete set of equations in
Sec.\ref{sec:obs} in order to facilitate comparisons.

Section \ref{sec:ses} describes the approaches taken by the SAID and MAID
groups in the determination of SE solutions. In the following section, Sec.\
\ref{sec:pseudodata}, we describe the event generator used to produce the
pseudo data employed in the SE and ED fits. The results of our fits are given
in a comparison of observables and multipole amplitudes in
Sec.\ref{sec:erange}. Our ED and SE solutions span a range of lab photon
energies $E_\gamma$ from 200 to 1200 MeV or total CM energies $W$ from 1120 to
1770 MeV. In order to limit the length of this article, we have included
figures for lab photon energies of 340 and 600 MeV corresponding to CM energies
of 1230 and 1420 MeV. The lower energy is in the region dominated by the
elastic $P_{33}$ partial wave. The higher energy is characterized generally by
the onset of larger inelasticities, except in the $P_{33}$ partial wave.
Section \ref{sec:conc} gives a discussion of the results and our conclusions.

\section{Amplitudes and Observables}
\label{sec:obs} In general, pion photoproduction from the nucleon, $\gamma N
\to \pi N$, is expressed by four invariant amplitudes, $A_i$, which are
covariant functions of two kinematical variables, such as the Mandelstam
variables $s,t$, or more often $E_\gamma,\theta$, the photon laboratory energy
and the pion CM angle. For given initial and final spin states $i,f$ the
transition matrix element is given by
\begin{equation}
\label{invari} t_{\pi\gamma}^{fi} = \bar u(p_f) \, \sum_{k=1}^{4}
\,A_k\, \varepsilon_\mu M_k^\mu\, u(p_i)
\end{equation}
where $u_{f,(i)}(p)$ is the Dirac spinor of the final (initial)
nucleon, normalized as $\bar u(p)u(p)=2
M_N$, $\varepsilon_\mu$ is the polarization vector of the photon, and
$M_k^\mu$ are a set of Dirac matrices defined to be consistent with
Ref.\cite{CGLN}. This leads to the exclusive photoproduction cross
section
\begin{equation}
\label{crossfi}
\frac{d\sigma^{f,i}}{d\Omega}=\frac{q}{k}
\left(\frac{M_N}{4\pi W}\right)^2 |t_{\pi\gamma}^{fi}|^2,
\end{equation}
where $q (k)$ is the magnitude of the three-momentum of the pion
(photon), $M_N$ is the mass of the nucleon, $W$ is the CM
scattering energy, and $t^{fi}_{\pi\gamma}$ is the transition matrix
element for pion photoproduction; the spin and internal quantum
numbers in this expression are implied. This matrix element is simply
expressed in terms of the Pauli spinors of the initial and final state
nucleons.

\begin{figure*}
\includegraphics[ width=300pt, keepaspectratio, angle=90]{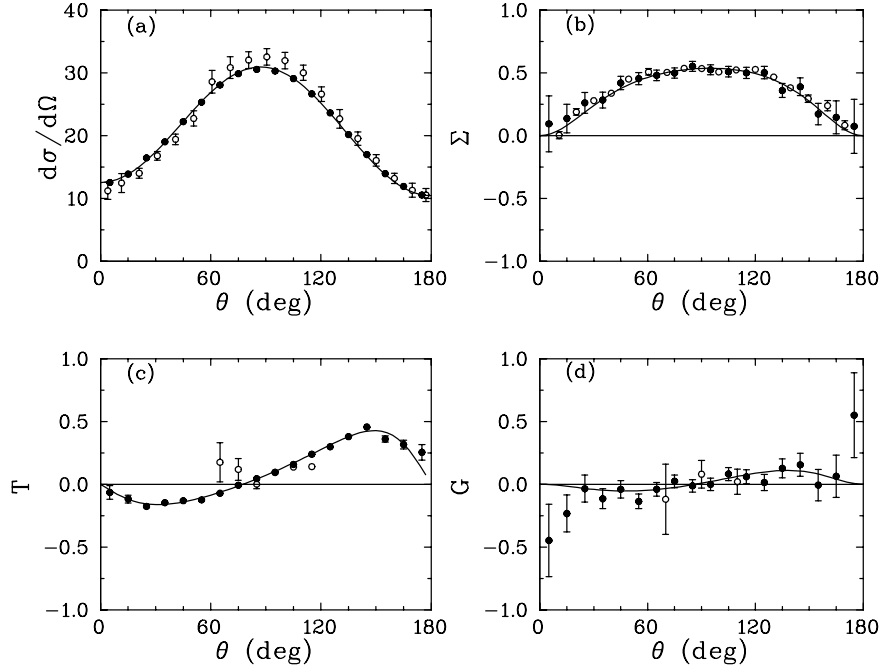}
\caption{\label{fig:pseudodata} Pseudo data (black circles) compared to
published experimental data (open circles) and the MAID ED solution MD07 (solid
curve) at photon beam energies of 320 MeV for $d\sigma / d\Omega$ (a), $\Sigma$
(b) and $T$ (c) and 340 MeV for the double spin observable $G$ (d).}
\end{figure*}

\subsection{CGLN and Helicity Amplitudes}
\label{subsec:cglnhel}

Expressed in terms of two-dimensional Pauli spinors, the matrix
element of the electromagnetic current takes the form
\begin{equation}\label{cgln1}
t^{fi}_{\pi\gamma} = - \frac{4\pi W}{M_N}\chi_f^\dagger {\cal F}
\chi_i,
\end{equation}
where $\chi_{f(i)}$ is a Pauli spinor for the nucleon in final (initial) state.
Following Ref.\cite{CGLN}, the operator ${\cal F}$ is decomposed into four
component amplitudes, ${ F}_i$, called the ``CGLN'' amplitudes, as
\begin{align}
\label{eq:cglnx}
{\cal F} &= -\epsilon_\mu J_{\pi N}^\mu \nonumber \\
         &= i\,({\vec {\sigma}}\cdot{ \hat{\epsilon}}) \, { F}_1
+ ({\vec {\sigma}} \cdot\hat { {q}})\, ({\vec
{\sigma}}\times\hat{{k}})\cdot{ \hat{\epsilon}}\, { F}_2 \nonumber \\
&+ i\,({\hat{\epsilon}}\cdot\hat { {q}})\, ({\vec {\sigma}} \cdot\hat {{k}})
  { F}_3
+ i ({ \hat{\epsilon}} \cdot \hat{{q}})({\vec {\sigma}}\cdot \hat { {q}}) \,
  { F}_4
\end{align}
where $\epsilon^\mu=(0,\vec{\epsilon})$ and $\vec{\epsilon}\cdot \vec{k} = 0$
for real photons. The multipole series of the CGLN amplitudes takes the
form\cite{CGLN,Pearl}:
\begin{align}
\label{eq:CGLN1}
{ F}_1  = \sum_{l=0}^\infty
\,[&(lM_{l+}+E_{l+})\,P_{l+1}^{'}(x) \nonumber \\
+ &((l+1)\,M_{l-}+E_{l-})\,P_{l-1}^{'}(x)] \,, \\
\label{eq:CGLN2}
{ F}_2  = \sum_{l=1}^\infty \,
[&(l+1)\,M_{l+}+lM_{l-}]\,P_l^{'}(x)\,,
\\
\label{eq:CGLN3}
{ F}_3 = \sum_{l=1}^\infty \, [&(E_{l+}-M_{l+})\,P_{l+1}^{''}(x) \nonumber \\
       + &(E_{l-}+M_{l-})\,P_{l-1}^{''}(x)]\,, \\
\label{eq:CGLN4}
{ F}_4  = \sum_{l=2}^\infty \,
[&M_{l+}-E_{l+}-M_{l-}-E_{l-}]\,P_l^{''}(x)\,,
\end{align}
where $x=\cos\theta$ is the cosine of the scattering angle.

The representation of the photoproduction amplitudes given by
Eqs.\eqref{eq:cglnx}--\eqref{eq:CGLN4} is useful in determining a
consistent notation since the CGLN amplitudes take this form in the
literature universally, to our knowledge. This provides a context for
the discussion of sign or phase conventions and various linear
combinations of amplitudes to form different bases. In this way, one
may readily discriminate various sign conventions used for
observables. While more compact expressions for the observables of
Eqs.\eqref{eqn:sig0}--\eqref{eqn:Tzp}, can be given in terms of the
helicity or transversity amplitudes (see Table \ref{tab:obs}), these
amplitudes are not uniquely defined. Differing phases and varying
linear combinations to define the helicity and transversity bases have
been used in the literature. In this work, we use the conventions
adopted by Walker\cite{Walker} for the helicity amplitudes. These
conventions are also used in Refs.\cite{Barker:1974vm,Barker75} of
Barker, Donnachie, Storrow (BDS). The relations in Table \ref{tab:obs}
give the reaction observables explicitly in terms of the helicity
amplitudes using the conventions of Ref.\cite{Walker}. Comparisons are
made there to works using differing sign conventions for convenience.

\begin{figure*}
\includegraphics[ width=300pt, keepaspectratio, angle=90]{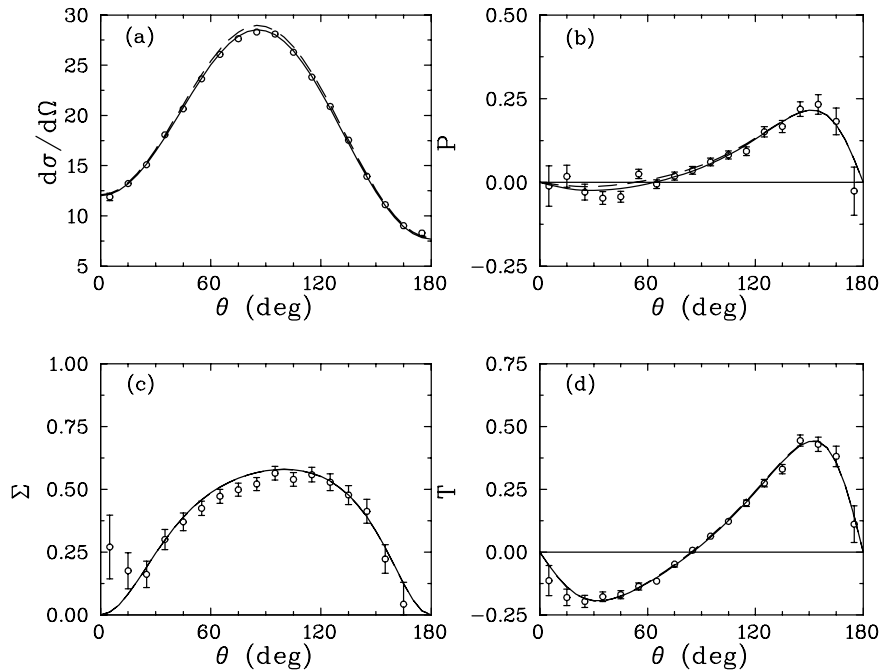}
\caption{\label{fig:f1}The SE4 fit (solid) and ED4 fit (dashed) 
compared to pseudo data (open circles) generated from MD07 for the $\pi^0 p$
channel for observables $d\sigma / d\Omega$ (a), $P$ (b), $\Sigma$ (c), and $T$
(d), at $E_\gamma=340$ MeV. Each figure includes two, usually overlapping,
curves.}
\end{figure*}

The helicity amplitudes are related to the CGLN amplitudes as
\begin{align}
H_1 &= -\frac{1}{\sqrt{2}}\,\mbox{sin}\,\theta\;
\mbox{cos}\,\frac{\theta}{2}\;({F}_3+{F}_4)\,,\\
H_2 &=
\sqrt{2}\,\mbox{cos}\,\frac{\theta}{2}\;[({F}_2-{F}_1)
 +\frac{1-\mbox{cos}\,\theta}{2}({F}_3-{F}_4)]\,,\\
H_3 &=  \frac{1}{\sqrt{2}}\,\mbox{sin}\,\theta\;
\mbox{sin}\,\frac{\theta}{2}\;({F}_3-{F}_4)\,,\\
H_4 &= \sqrt{2}\,\mbox{sin}\,\frac{\theta}{2}\;[({F}_1+{F}_2)
 +\frac{1+\mbox{cos}\,\theta}{2}({F}_3+{F}_4)]\, ,
\end{align}
which are identical to Eqs.(24) of Ref.\cite{Walker}.  The outgoing pion
direction is specified by the scattering angle $\theta$ and azimuthal angle
$\phi$. We have chosen $\phi=0$ by aligning the coordinate system with the
reaction plane as in Fig.\eqref{fig:frames} (see the discussion immediately
following Eq.\eqref{eqn:TR} below).  The same helicity amplitudes are
employed by BDS\cite{Barker:1974vm,Barker75} with, however, a distinct naming
convention:
\begin{align}
\label{eqn:BDS2Walker}
S_1 &= H_1, & N   &= H_2,\nonumber \\
D   &= H_3, & S_2 &= H_4.
\end{align}

\subsection{Coordinate Frames and Polarizations}
\label{subsec:frames}

Experiments with three types of polarization may be performed in meson
photoproduction: photon beam polarization, polarization of the target
nucleon, and polarization of the recoil nucleon. The target
polarization is described in the frame $\{ x, y, z \}$ of
Fig.\eqref{fig:frames}, with the $z$-axis coincident with the photon
momentum $\hat{ \vec k}$, the $y$-axis perpendicular to the reaction
plane, ${\hat{ \vec y}} = {\hat{ \vec k}} \times {\hat{ \vec q}} /
\sin \theta$, and the $x$-axis given by ${\hat{\vec x}} = {\hat{ \vec
y}} \times {\hat{\vec z}}$. For recoil polarization we use the frame
$\{ x', y', z' \}$, with the $z'$-axis coincident with the momentum of
the outgoing meson ${\hat{\vec q}}$, the $y'$-axis coincident with the
$y$-axis, as for the target polarization, and the $x'$-axis given by
${\hat{\vec{x}}'} = {\hat{\vec{y}}'} \times {\hat{\vec{z}}'}$.

The photon polarization is either linear or circular. A linearly polarized
photon is defined by the degree of transverse polarization, $P_T$, and the
angle $\varphi$ of the polarization plane relative to the reaction plane or,
equivalently, to the $\hat{\vec x}$ direction. For example, a beam completely
polarized in the reaction plane has $\varphi=0$ (with $P_T=1$); for a beam
polarized perpendicular to the reaction plane, ${\hat{ \vec y}}$, the
polarization angle is $\varphi=\pi/2$ (with $P_T=1$). For a photon of
right-(left-)circular polarization $P_{\odot}=\pm 1$.

The target nucleon polarization is specified by three polarization
components $(P_x,P_y,P_z)$ with respect to the $(\hat{\vec
x},\hat{\vec y},\hat{\vec z})$ coordinate system, displayed in
Fig.\eqref{fig:frames}. The recoil nucleon polarization is specified
similarly by $(P_{x'},P_{y'},P_{z'})$ with respect to $(\hat{\vec
x}',\hat{\vec y}',\hat{\vec z}')$ also shown in
Fig.\eqref{fig:frames}.

\begin{figure*}
\includegraphics[ width=300pt, keepaspectratio, angle=90]{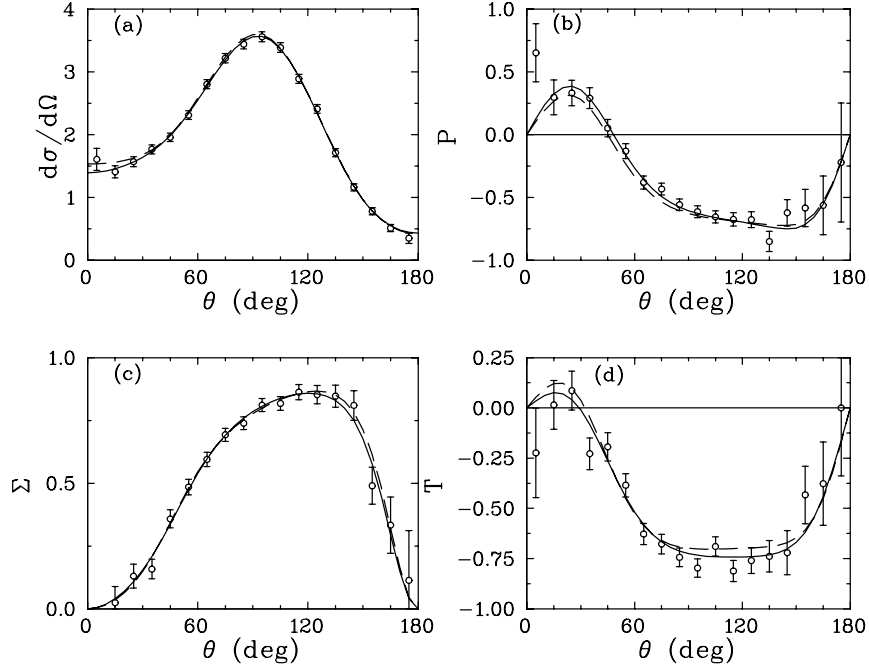}
\caption{\label{fig:f2}The SE4 fit (solid) and ED4 (dashed) compared to pseudo
data (open circles) for $\pi^0 p$ at 600 MeV, as in Fig.\eqref{fig:f1}.}
\end{figure*}

\subsection{Polarization Observables}
\label{subsec:polobs} We may classify the differential cross section for
general polarization states of the beam, target, and recoil particles by three
classes of double polarization experiments.
Using the notation described in the
previous subsection, measurements with polarized photons and a polarized
target, the \textit{beam-target} experiments, are given by the differential
cross section
\begin{align}
\label{eqn:BT}
\frac{1}{\sigma_0}
\frac{d \sigma}{d \Omega} & =
1 - P_T \Sigma \cos 2 \varphi \nonumber \\
& + P_x \left( - P_T H \sin 2 \varphi + P_{\odot} F \right)
\nonumber \\
& + P_y \left( T - P_T P \cos 2 \varphi \right) \nonumber \\
& + P_z \left( P_T G \sin 2 \varphi - P_{\odot} E \right) ;
\end{align}
for polarized photons and measured recoil polarization,
\textit{beam-recoil} experiments, we obtain
\begin{align}
\label{eqn:BR}
\frac{1}{\sigma_0}
\frac{d \sigma}{d \Omega} & =
1 - P_T \Sigma \cos 2 \varphi \nonumber \\
& + P_{x'} \left( -P_T O_{x'} \sin 2 \varphi - P_{\odot} C_{x'}
\right) \nonumber \\
& + P_{y'} \left( P - P_T T \cos 2 \varphi \right) \nonumber \\
& + P_{z'} \left( -P_T O_{z'} \sin 2 \varphi  - P_{\odot}
C_{z'} \right) ;
\end{align}
for the polarized target and recoil polarization measurements,
\textit{target-recoil} experiments:
\begin{align}
\label{eqn:TR}
\frac{1}{\sigma_0}
\frac{d \sigma}{d \Omega} & = 1 + P_{y} T +
P_{y'} P
+ P_{x'} \left( P_x T_{x'} - P_{z} L_{x'} \right) \nonumber \\
& + P_{y'} P_y \Sigma  + P_{z'}\left( P_x T_{z'} + P_{z} L_{z'}\right) .
\end{align}
In these equations, $\sigma_0$ denotes the unpolarized differential cross
section. Here $\varphi$ is the azimuthal angle of the photon polarization
vector with respect to the reaction plane. Alternatively, one could fix the
photon polarization vector and observe the pion out of the polarization plane
of the photon. In this case one would obtain a pion angle $\phi=-\varphi$. This
leads to a minus sign in all terms proportional to $\sin 2 \varphi$. Equations
\eqref{eqn:BT}, \eqref{eqn:BR}, and \eqref{eqn:TR} are identical with Eqs.(2),
(3), and (4) of Ref.\cite{Barker75} upon conversion of their $\sigma_i \to
P_{i'}$, $O_i \to O_{i'}$, $C_i \to C_{i'}$, $T_i \to T_{i'}$, and $L_i \to
L_{i'}$. Note, however, that our coordinate frames are identical to those of
BDS. The triple polarization cross section, which specifies the polarizations
of the beam, target, and recoil nucleon simultaneously, provides no additional
information in the pion photoproduction reaction.  

\begin{figure*}
\includegraphics[ width=300pt, keepaspectratio, angle=90]{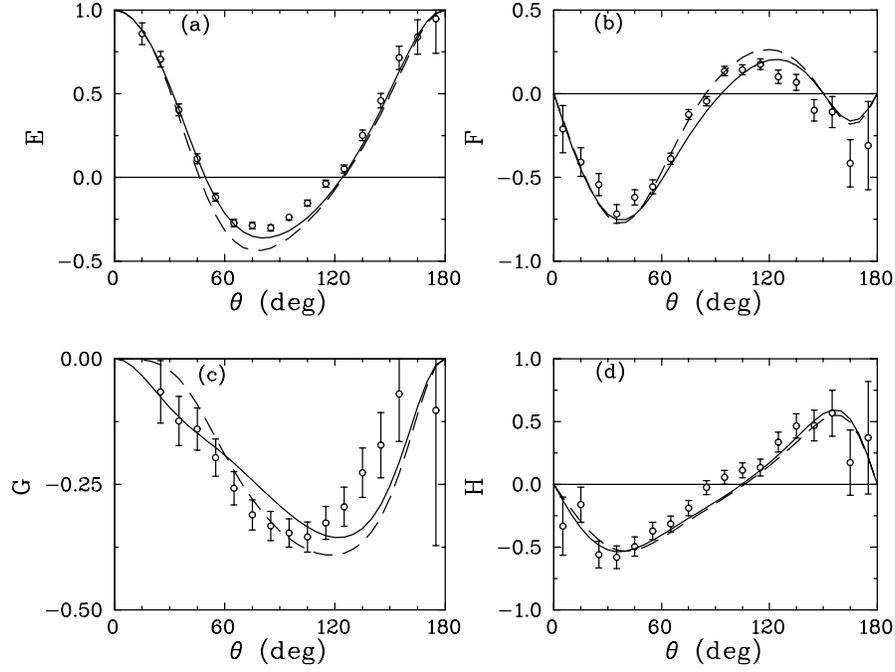}
\caption{\label{fig:f3}The SE4 fit (solid) and ED4 (dashed) compared to pseudo
data (open circles) for $\pi^0 p$ beam-target observables, $E$ (a), $F$ (b),
$G$ (c), and $H$ (d) at 600 MeV.}
\end{figure*}

The spin observables, expressed in terms of the CGLN amplitudes,
within the sign convention of Barker, Donnachie and Storrow,
have been consistently employed in both the MAID and SAID
parametrizations. They are given explicitly as:
\begin{widetext}
\begin{align}
\label{eqn:sig0}
\sigma_{0}   & =  \,\mbox{Re}\,\left\{ \fpf{1}{1} + \fpf{2}{2} +
\sin^{2}\theta\,(\fpf{3}{3}/2
+ \fpf{4}{4}/2 + \fpf{2}{3} + \fpf{1}{4}
+ \cos\theta\,\fpf{3}{4}) - 2\cos\theta\,\fpf{1}{2} \right\} \rho \\
\hat{\Sigma} & = -\sin^{2}\theta\;\mbox{Re}\,\left\{
\left(\fpf{3}{3} +\fpf{4}{4}\right)/2
+ \fpf{2}{3} + \fpf{1}{4} + \cos\theta\,\fpf{3}{4}\right\}\rho \\
\hat{T}      & =  \sin\theta\;\mbox{Im}\,\left\{\fpf{1}{3} -
\fpf{2}{4} + \cos\theta\,(\fpf{1}{4} - \fpf{2}{3})
                   - \sin^{2}\theta\,\fpf{3}{4}\right\}\rho \\
\hat{P}      & = -\sin\theta\;\mbox{Im}\,\left\{ 2\fpf{1}{2} +
\fpf{1}{3} - \fpf{2}{4} - \cos\theta\,(\fpf{2}{3} -\fpf{1}{4})
                   - \sin^{2}\theta\,\fpf{3}{4}\right\}\rho \\
\hat{E}      & =  \,\mbox{Re}\,\left\{ \fpf{1}{1} + \fpf{2}{2} -
2\cos\theta\,\fpf{1}{2}
                   + \sin^{2}\theta\,(\fpf{2}{3} + \fpf{1}{4}) \right\}\rho \\
\hat{F}      & =  \sin\theta\;\mbox{Re}\,\left\{\fpf{1}{3} - \fpf{2}{4} - \cos\theta\,(\fpf{2}{3} - \fpf{1}{4})\right\}\rho \\
\hat{G}      & =  \sin^{2}\theta\;\mbox{Im}\,\left\{\fpf{2}{3} + \fpf{1}{4}\right\}\rho \\
\hat{H}      & =  \sin\theta\;\mbox{Im}\,\left\{2\fpf{1}{2} +
\fpf{1}{3} - \fpf{2}{4}
                   + \cos\theta\,(\fpf{1}{4} - \fpf{2}{3})\right\}\rho \\
\hat{C}_{x'} & =  \sin\theta\;\mbox{Re}\,\left\{\fpf{1}{1} -
\fpf{2}{2} - \fpf{2}{3} + \fpf{1}{4}
                   - \cos\theta\,(\fpf{2}{4} - \fpf{1}{3})\right\}\rho \\
\hat{C}_{z'} & = \,\mbox{Re}\,\left\{2\fpf{1}{2} -
\cos\theta\,(\fpf{1}{1} + \fpf{2}{2})
                   + \sin^{2}\theta\,(\fpf{1}{3} + \fpf{2}{4})\right\}\rho \\
\hat{O}_{x'} & = \sin\theta\;\mbox{Im}\,\left\{\fpf{2}{3} - \fpf{1}{4} + \cos\theta\,(\fpf{2}{4} - \fpf{1}{3})\right\}\rho \\
\hat{O}_{z'} & = - \sin^{2}\theta\;\mbox{Im}\,\left\{\fpf{1}{3} + \fpf{2}{4}\right\}\rho\\
\hat{L}_{x'} & = - \sin\theta\;\mbox{Re}\,\left\{\fpf{1}{1} -
\fpf{2}{2} - \fpf{2}{3} + \fpf{1}{4}
+ \sin^{2}\theta\,(\fpf{4}{4} - \fpf{3}{3})/2
+ \cos\theta\,(\fpf{1}{3} - \fpf{2}{4})\right\}\rho \\
\hat{L}_{z'} & =  \,\mbox{Re}\,\left\{2\fpf{1}{2} -
\cos\theta\,(\fpf{1}{1} + \fpf{2}{2})
+ \sin^{2}\theta\,(\fpf{1}{3} + \fpf{2}{4} + \fpf{3}{4})
+ \cos\theta\sin^{2}\theta\,(\fpf{3}{3} + \fpf{4}{4})/2 \right\}\rho \\
\hat{T}_{x'} & = -\sin^{2}\theta\;\mbox{Re}\,\left\{\fpf{1}{3} +
\fpf{2}{4} + \fpf{3}{4}
+ \cos\theta\,(\fpf{3}{3} + \fpf{4}{4})/2 \right\}\rho \\
\label{eqn:Tzp}
\hat{T}_{z'} & =  \sin\theta \;\mbox{Re}\, \left\{\fpf{1}{4} - \fpf{2}{3}
+ \cos\theta\,(\fpf{1}{3} - \fpf{2}{4})
+ \sin^{2}\theta\,(\fpf{4}{4} - \fpf{3}{3})/2 \right\}\rho,
\end{align}
\end{widetext}
where $\hat{\Sigma}={\Sigma}\,\sigma_0,$ etc. and $\rho=q/k$.

\begin{table*}
\begin{center}
\begin{tabular}{|c|c|c|c|}
\hline
 spin          &  helicity                                             & comparison with &  comparison with   \\
 observables   &  representation                                       & Fasano et al.~\cite{FTS92}  & Chiang et al.~\cite{chiang}                                         \\
\hline
$\sigma_0$     & $\frac{1}{2}(|H_1|^2 + |H_2|^2 + |H_3|^2 + |H_4|^2)$  & $+\sigma_0$     & $+\sigma_0       =+\check{\Omega}_{1} $  \\
$\hat{\Sigma}$ & Re$(H_1 H_4^* - H_2 H_3^*)$                           & $+\hat{\Sigma}$ & $+\hat{\Sigma}   =+\check{\Omega}_{4} $  \\
$\hat{T}$      & Im$(H_1 H_2^* + H_3 H_4^*)$                           & $+\hat{T}$      & $-\hat{T}        =+\check{\Omega}_{10}$  \\
$\hat{P}$      & $-$Im$(H_1 H_3^* + H_2 H_4^*)$                        & $+\hat{P}$      & $+\hat{P}        =+\check{\Omega}_{12}$  \\
\hline
$\hat{G}$      & $-$Im$(H_1 H_4^* + H_2 H_3^*)$                        & $+\hat{G}$      & $-\hat{G}        =-\check{\Omega}_{3} $  \\
$\hat{H}$      & $-$Im$(H_1 H_3^* - H_2 H_4^*)$                        & $-\hat{H}$      & $-\hat{H}        =-\check{\Omega}_{5} $  \\
$\hat{E}$      & $\frac{1}{2}(-|H_1|^2 + |H_2|^2 - |H_3|^2 + |H_4|^2)$ & $-\hat{E}$      & $-\hat{E}        =-\check{\Omega}_{9} $  \\
$\hat{F}$      & Re$(H_1 H_2^* + H_3 H_4^*)$                           & $+\hat{F}$      & $-\hat{F}        =-\check{\Omega}_{11}$  \\
\hline
$\hat{O_{x'}}$ & $-$Im$(H_1 H_2^* - H_3 H_4^*)$                        & $-\hat{O_{x'}}$ & $-\hat{O_{x'}}   =-\check{\Omega}_{14}$  \\
$\hat{O_{z'}}$ & Im$(H_1 H_4^* - H_2 H_3^*)$                           & $-\hat{O_{z'}}$ & $-\hat{O_{z'}}   =+\check{\Omega}_{7} $  \\
$\hat{C_{x'}}$ & $-$Re$(H_1 H_3^* + H_2 H_4^*)$                        & $-\hat{C_{x'}}$ & $+\hat{C_{x'}}   =-\check{\Omega}_{16}$  \\
$\hat{C_{z'}}$ & $\frac{1}{2}(-|H_1|^2 - |H_2|^2 + |H_3|^2 + |H_4|^2)$ & $-\hat{C_{z'}}$ & $+\hat{C_{z'}}   =-\check{\Omega}_{2} $  \\
\hline
$\hat{T_{x'}}$ & Re$(H_1 H_4^* + H_2 H_3^*)$                           & $+\hat{T_{x'}}$ & $+\hat{T_{x'}}   =-\check{\Omega}_{6} $  \\
$\hat{T_{z'}}$ & Re$(H_1 H_2^* - H_3 H_4^*)$                           & $+\hat{T_{z'}}$ & $+\hat{T_{z'}}   =-\check{\Omega}_{13}$  \\
$\hat{L_{x'}}$ & $-$Re$(H_1 H_3^* - H_2 H_4^*)$                        & $-\hat{L_{x'}}$ & $+\hat{L_{x'}}   =+\check{\Omega}_{8} $  \\
$\hat{L_{z'}}$ & $\frac{1}{2}(|H_1|^2 - |H_2|^2 - |H_3|^2 + |H_4|^2)$  & $+\hat{L_{z'}}$ & $-\hat{L_{z'}}   =-\check{\Omega}_{15}$  \\
\hline
\end{tabular}
\caption{\label{tab:obs}Spin observables expressed by helicity
amplitudes in the notation of Walker~\cite{Walker}. The sign
definition is taken from Barker, Donnachie and Storrow~\cite{Barker75}
by replacing $N\rightarrow H_2,\; S_1\rightarrow H_1,\; S_2\rightarrow
H_4,\; D\rightarrow H_3$ as given in the text. This sign definition is
used by SAID and MAID. In column three, we compare the sign
definitions of Fasano, Tabakin and Saghai~\cite{FTS92}, which is also
used recently by Refs.~\cite{Anisovich:2009zy,Dey:2010fb}.
Furthermore we also give the sign definition of Chiang and
Tabakin~\cite{chiang} and the relations to the $\Omega$ observables,
that are the basis of the Fierz consistency relations given in
Ref.~\cite{chiang}.}
\end{center}
\end{table*}

\section{Single-energy solutions}
\label{sec:ses}

The SE solutions afford the most faithful description of the data, in
terms of the lowest possible $\chi$-squared, consistent with constraints from the ED
solutions. They have been used to probe the data for structure that may
not have been properly encoded by the smooth ED solution.  Both MAID
and SAID SE solutions are determined by constrained fits to the data
in sufficiently narrow energy bins. These constraints limit the
variability in the SE results through the assumption that the global ED fit
is close to the `true' solution. This assumption appears to be
validated by the fact that the $\chi$-squared of both the ED and SE
solutions suggest a realistic description of the data. Since the ED
results are used to constrain the SE solutions in different ways
within the MAID and SAID parametrization approaches, we begin this
section with a brief description of their respective methods.

\subsection{SAID SE Solutions}
\label{subsec:saidses} The SAID SE solutions are obtained in bins of the
scattering energy spanning a few MeV. Starting values for the SE solutions are
given by the multipole moduli and phases determined by ED solutions. In order
to account for the variation of the modulus and phase over the width of an
energy bin, a linear approximation to the energy dependence for each quantity
is taken from the ED fit. The bin width increases from 5 to 20 MeV as the
energy increases into regions with sparse measurements. Only the central values
of the modulus and phase are searched for each SE solution; the slopes are held
fixed. For the purposes of the present study, however, the pseudo data are
generated at single energies coinciding with the central energy value of the
bin. This removes the need to estimate the energy variation based on an ED fit.
In order to constrain the SE multipole solutions to be close to the ED values,
the ED multipoles themselves are ascribed arbitrary errors and fitted along
with experimentally determined reaction data of the bin under consideration. In
general, these multipole `pseudo data' contribute a negligible amount to the
resultant chi-squared.

In all previously published sets of SE fits, the multipole phases were
fixed to ED values when fitting to the data.  As a result, multipole
phases would change only through a modification of the ED fit.  The
number of searched multipoles has generally been subjective but may be
increased until a $\chi^2$/data near unity is approached.

As previously mentioned, the SAID SE fits were generated originally to
search for structures not seen in the ED fit, and they have been used to
re-initialize the ED fit, with the hope that this procedure would converge
to the correct solution. As described above, they are not independent
of the ED fit.  It should be emphasized that the ED and SE solutions
are fits to the experimentally observed reaction data; the ED solution
is \textit{not} a fit to the SE values.

\subsection{MAID SE solutions}
\label{subsec:maidses} The MAID partial wave analysis follows a similar
two-step approach\cite{Drechsel:2007if}. In the first step, a global ED
solution is determined first by fitting all experimentally observed reaction
data, in a similar fashion to the SAID ED approach, in the range 140~MeV$\le
E_{\gamma} \le$1610~MeV. This allows the determination of the phases of each
multipole, i.e., the ratio $ Im\, t_{\pi\gamma}^{\alpha}/Re\,
t_{\pi\gamma}^{\alpha}$, where $\alpha$ is the multipole, above the two-pion
threshold. At energies below the two-pion threshold, this phase is constrained,
by Watson's theorem, to be equal to the $\pi N$ scattering phase. In the second
step we perform local SE fits to the data in energy bins of 10 MeV (over the
range 140~MeV$\le E_{\gamma} \le$460~MeV) and 20 MeV (for the higher energies)
by varying the absolute values of the multipoles but keeping the phase fixed.
In order to damp strong local variations we introduce a penalty factor similar
to a Bayesian approach and minimize the modified $\chi^2$ function
\begin{eqnarray}
\chi^2 = \sum_i^{N_{data}}
\left(\frac{\Theta_i-\Theta_i^{exp}}{\delta\Theta_i}\right)^2 +
\sum_j^{N_{mult}} \left(\frac{X_j-1}{\Delta}\right)^2\,.
\label{chi2}
\end{eqnarray}
The first term on the right-hand side of this equation is the standard $\chi^2$
function with $\Theta_i$ the calculated and $\Theta_i^{exp}$ the measured
observables, $\delta\Theta_i$ the statistical errors, and $N_{data}$ the number
of data points. In the second term, $N_{mult}$ is the number of the varied
multipoles and $X_j$ is the fitting parameter describing the deviation from the
global fit, that is, the fitting procedure starts with the initial value
$X_j=1$ corresponding to the global solution, for each multipole indexed by
$j$. The quantity $\Delta$ enforces a smooth energy dependence of the single
energy solution. In the limit of $\Delta\rightarrow\infty$ we obtain the
standard $\chi$-squared, $\chi^2_{std}$, and for $\Delta\rightarrow 0$ the
single energy and the global solutions become identical. The optimum value for
$\Delta$ is chosen from the condition $1<\chi^2/\chi^2_{std}<1.05$. The
described two-step fitting procedure may be repeated several times by adjusting
the energy dependence of the global solution, for example by changing the
resonance parametrization to obtain a better agreement between the global and
local solutions.

\section{Generating pseudo data}
\label{sec:pseudodata}
Measurements of the spin observables defined in Sec.
\ref{subsec:polobs}, over a large angular range with accuracy
sufficient to have an impact on multipole analyses, require
experiments of high intensity, with linearly and circularly polarized
photon beams, strongly polarized targets and accurate recoil
polarimetry together with hermetic detector systems. Such experiments
have become technically possible only very recently at the tagged
photon facilities at ELSA (Univsersity of Bonn), CEBAF (Jefferson
Laboratory) and MAMI (University of Mainz).

\begin{table*}
\begin{tabular*}{0.75\textwidth}{@{\extracolsep{\fill}}cccc}
Multipole & SE4 & MD07 & SP09  \\
\hline
$E_{0+}^{1/2}$ & 9.40(0.08) &   9.36 & 7.30 \\
$E_{0+}^{3/2}$ &  18.06(0.16) & 17.91 & 15.87 \\
$M_{1-}^{1/2}$ & 2.28(0.15) &   2.21 & 1.65 \\
$M_{1-}^{3/2}$ &  9.26(0.27) & 9.31 & 7.89  \\
$E_{1+}^{1/2}$ &  1.82(0.05) & 1.79 & 1.76 \\
$M_{1+}^{1/2}$ &  2.52(0.18) & 2.55 & 1.99 \\
$E_{1+}^{3/2}$ &  1.00(0.07) & 1.12 & 1.08 \\
$M_{1+}^{3/2}$ & 51.91(0.06) & 52.0 & 55.25 \\
\end{tabular*}
\caption{\label{tab:pole3}Single-energy fit (SE4) to MD07 pseudo data compared
to MD07 ED values at 340 MeV. The SP09 solution, fitted to the SAID database,
is displayed for comparison. Multipoles given in millifermi units.}
\end{table*}

For this study, we have generated pseudo data for single and double
polarization observables of the $\gamma p \to \pi^0 p$ and $\gamma p \to \pi^+
n$ reactions with statistical uncertainties comparable to those expected at the
above mentioned precision electromagnetic facilities within the next few years
for energies from the reaction thresholds up to a photon beam energy of
$E_{\gamma} = 1500$~MeV ($W=1920$~MeV)

Events for the neutral and charged pion reactions are generated via a Monte
Carlo algorithm. A sample of reaction parameters -- the beam energy, meson
scattering angles ($\theta$ and $\phi$), circular and linear beam
polarizations, longitudinal and transverse target polarizations and the recoil
nucleon spin alignment angles ($\theta_R$ and $\phi_R$) -- are drawn from a
weight function given by the polarized cross section,
Eq.\eqref{eqn:BT}-\eqref{eqn:TR}. The events were generated for beam energy
bins of $\Delta E_{\gamma}=10$~MeV and angular bins of
$\Delta\theta_\pi^{cm}=10^\circ$, based on the MAID (MD07) model predictions.
Beam polarizations are assumed to be $P_T = 60\%$ for linearly polarized
photons and $P_\odot = 70\%$ for circularly polarized photons. These are
typical values resulting from $e^-$ to $\gamma$ helicity transfer (giving
circular photon polarization) and coherent bremsstrahlung (giving linear photon
polarization), which are achieved at tagged photon facilities. We assume that
the beam polarizations have no energy dependence and therefore should be
interpreted as resulting from various sets of experimental conditions. For
target protons an average polarization of $P=80\%$ is assumed, as is available
with butanol frozen spin targets \cite{poltarg}.  The polarization of recoiling
nucleons may be measured using a subsequent scattering reaction in a carbon
analyzer.  For these measurements an average analyzing power of $A=20\%$ is
assumed, close to values typically achievable. For each observable, typically
$5\cdot10^6$ events have been generated over the full energy range. The pseudo
data have not been folded with any additional acceptance given by a specific
detector system.

The pseudo data are an accurate representation of both the MAID MD07 ED
solution and the observed reaction data, as shown in
Fig.\eqref{fig:pseudodata}. Here, the unpolarized cross section,
$d\sigma/d\Omega$, the photon beam asymmetry, $\Sigma$, the target asymmetry,
$T$, and the beam-target observable, $G$ are shown at energies corresponding to
the $\Delta (1232)$ resonance.

In the following section, we discuss the application of the SAID
parametrization approach to obtain ED and SE solutions in fits to the pseudo
data. We compare these solutions to the MAID MD07 ED solution and the pseudo
data and discuss various features of their energy dependence.

\section{Dependence on Energy}
\label{sec:erange} Using the pseudo data generated from the MAID MD07 ED
solution we apply the existing SAID parametrization
approach\cite{Arndt:1989ww,Arndt:2002wr} to obtain the pion photoproduction
multipoles.  The results given below cover values of $E_{\gamma}$ from 200 MeV
to 1.2 GeV. The determination of a set of isospin-1/2 and 3/2 multipoles in the
low-energy region is problematic due to the distinct thresholds for $\pi^0 p $
and $\pi^+ n$ photoproduction from the proton. We avoid this issue by fitting
data above the $\pi^+ n$ threshold.

\begin{figure*}
\includegraphics[ width=300pt, keepaspectratio, angle=90]{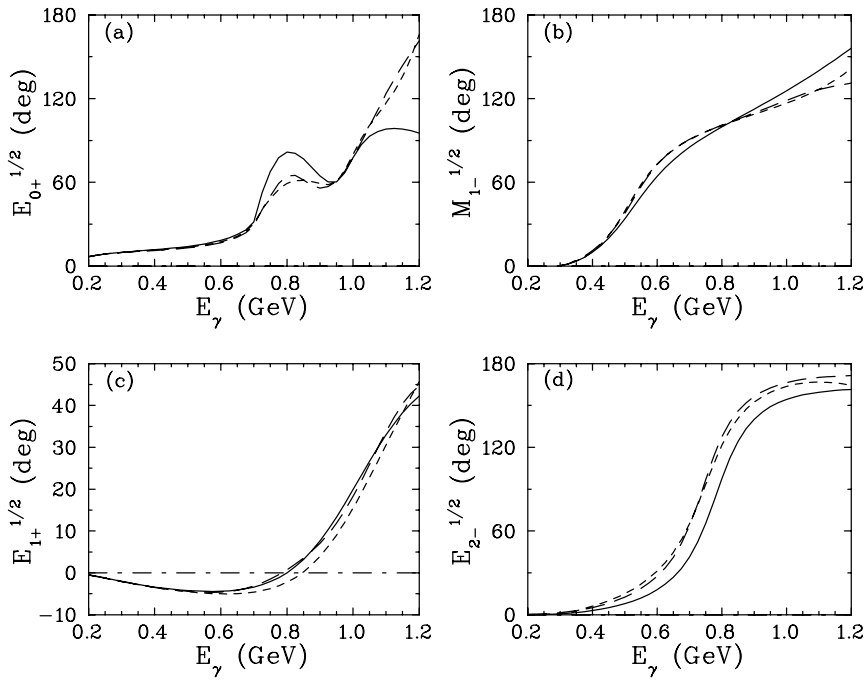}
\caption{\label{fig:f4} Multipole phases from energy-dependent fits.
SP09 (solid), MD07 (dashed), ED4 (dotted).}
\end{figure*}

The use of Watson's theorem, greatly simplifies SE fits at energies where it is
valid (depending on the multipole, the Watson regime is typically somewhat
higher than the first inelastic, two-pion threshold).  Here, the multipole
phase is fixed to the corresponding $\pi N$ elastic partial wave phase and only
the modulus allowed to vary. 
The SAID ED fit form, for each multipole, is
\begin{equation}
M = \alpha (1+ i T_{\pi N} ) + \beta T_{\pi N} + \gamma ({\rm Im} T_{\pi N}
- |T_{\pi N}|^2 ),
\end{equation}
$T_{\pi N}$ being the associated $\pi N$ partial-wave amplitude,
which allows a smooth transition from the Watson regime. Here $\alpha$ contains
the Born contributions plus a (real) phenomenological term, $\beta$ (also real)
is purely phenomenological, and $\gamma$ is a complex polynomial allowing
a further departure from the Watson regime, proportional to the $\pi N$
reaction cross section. 

As shown in Table \ref{tab:pole3}, the SAID SE fit
to MAID pseudo data, at the $\Delta(1232)$ resonance energy ($E_\gamma=340$
MeV), reproduces the dominant amplitudes. The SE result is, however,
significantly different from an ED fit (SP09)\cite{SAID-unp} to the full SAID
database of experimental observables, covering the resonance region. We note
that this database, determining the SP09 solution, includes few
measured polarization observables.  In producing the SE result, SE4, only 
$d\sigma/d\Omega$, $P$, $\Sigma$, and $T$ have been fitted.

\begin{table*}
\begin{tabular*}{0.75\textwidth}{@{\extracolsep{\fill}}ccccccc}
Multipole & $\pi N$ PW & MD07 & SP09 & ED4 & SE4p & SE8p \\
\hline
$E_{0+}^{1/2}$ & $S_{11}$ & 16.7 & 18.4  & 17.4  & 16.0(3.0)  & 16.2(0.9) \\
$M_{1-}^{1/2}$ & $P_{11}$ & 72.7 & 64.4  & 73.2  & 68.2(2.4)  & 73.4(1.6) \\
$M_{1-}^{3/2}$ & $P_{31}$ & 163.9 & 167.1 & 172.0 & 176.8(6.5) & 167.5(1.8)\\
$E_{2-}^{1/2}$ & $D_{13}$ & 27.6 & 17.9  & 31.4  & 27.5(3.1)  & 26.1(1.0) \\
$M_{2-}^{1/2}$ & $D_{13}$ & 26.8 & 22.2  & 26.7  & 25.5(1.8)  & 26.7(1.0) \\
\end{tabular*}
\caption{\label{tab:pole6}Multipole phases (degrees) from
single-energy fits to 4 (SE4p) and 8 (SE8p)
observables at 600 MeV (phases searched), compared to the energy-dependent
fits SP09, ED4, and MD07 (see text). Also listed is the associated
$\pi N$ partial wave (PW).}
\end{table*}

\begin{figure*}
\includegraphics[width=0.35\textwidth, keepaspectratio, angle=90]{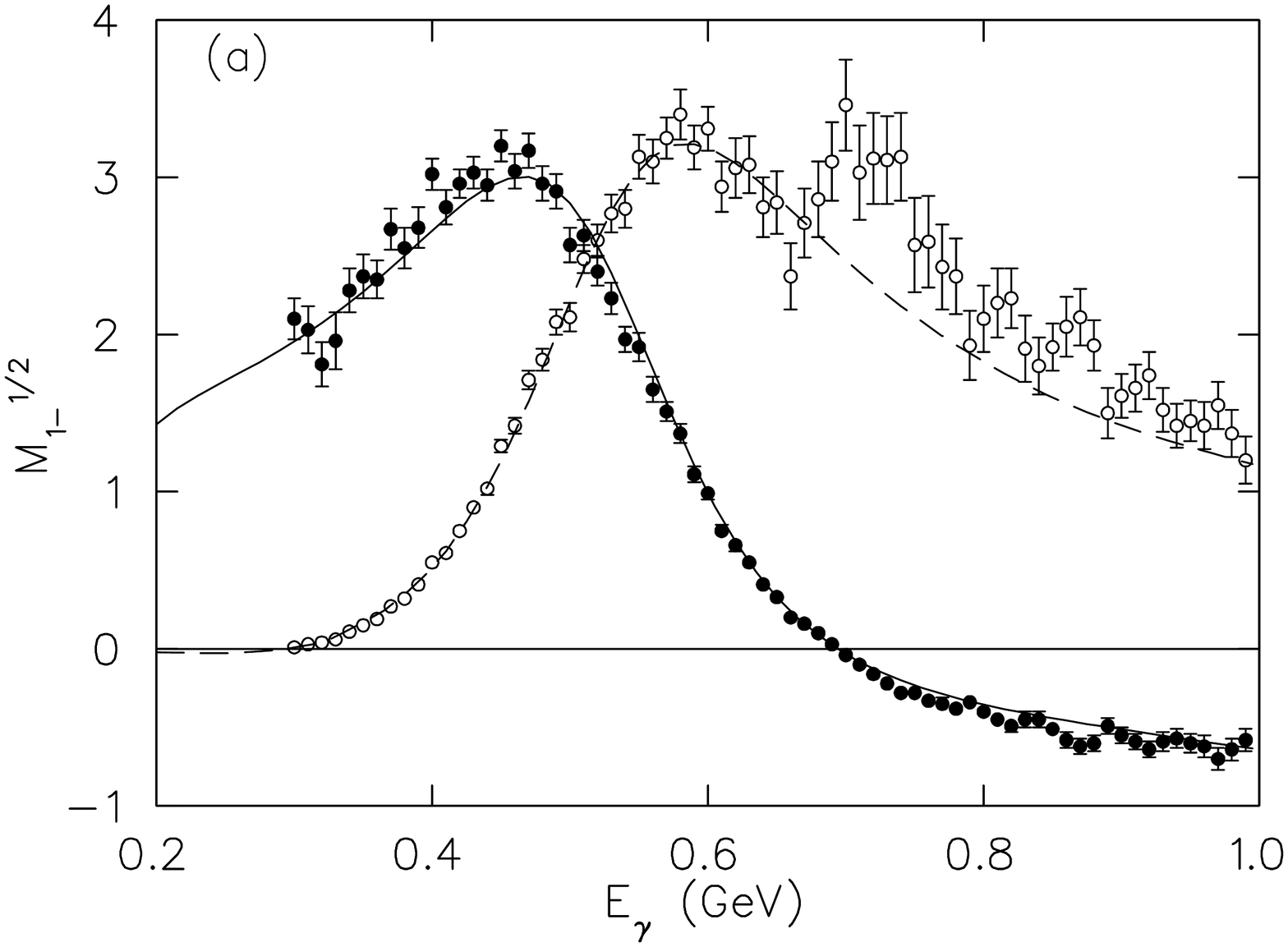}
\hspace*{0.2cm}
\includegraphics[width=0.35\textwidth, keepaspectratio, angle=90]{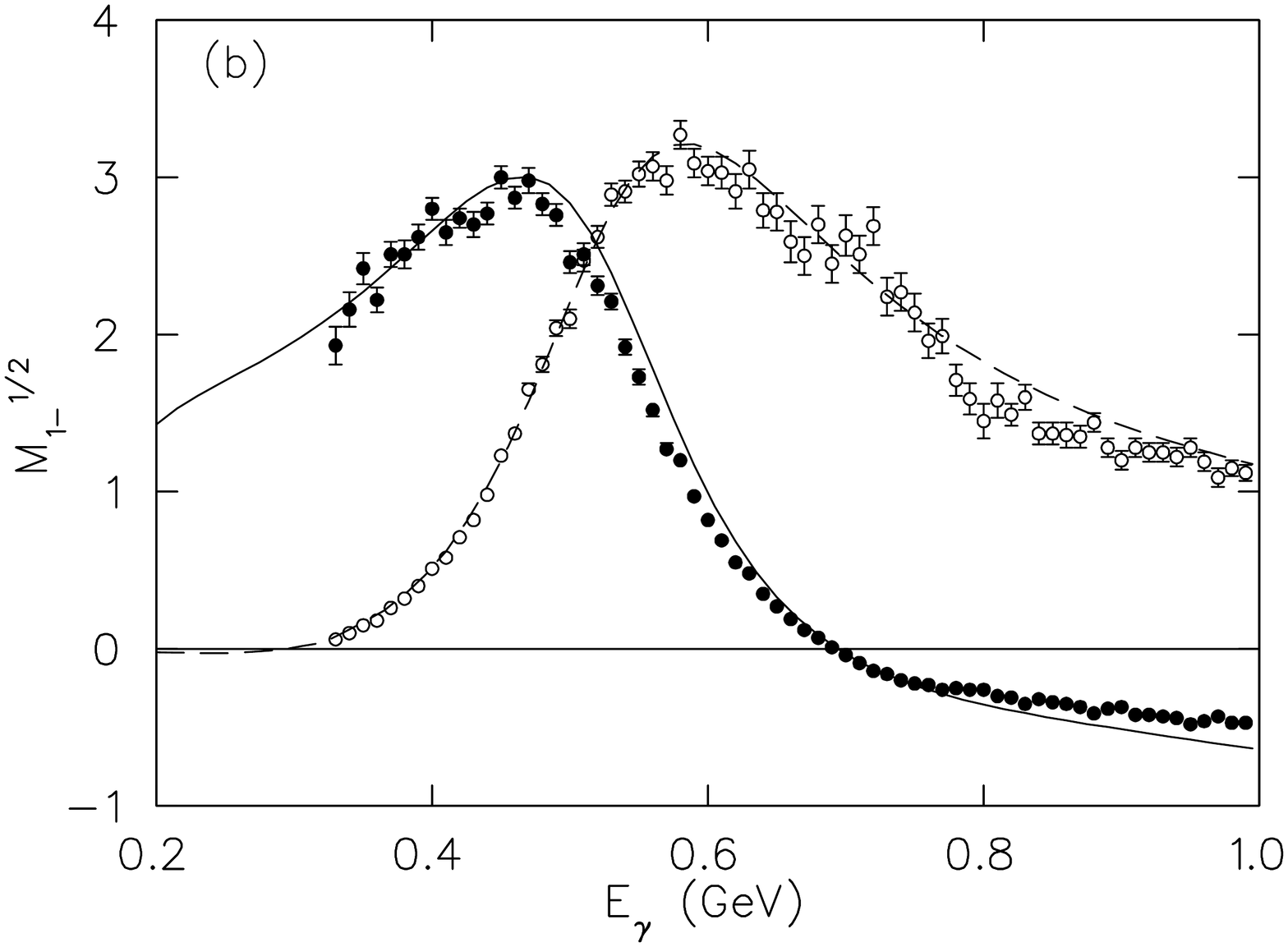}
\caption{\label{fig:pwp11} Real and imaginary parts of the $P_{11}(pM)$ partial
wave amplitude ($M_{1-}^{1/2}$). The solid (dashed) line shows the real
(imaginary) part of the MD07 solution, used for the pseudo data generation. (a)
Solid (filled) points display the SE4 fits to $d\sigma/d\Omega$
and the three single-spin observables $\Sigma$, $T$ and $P$. (b)
Notation as in (a) for SE8 fits also including the beam-target spin observables
$E$, $F$, $G$ and $H$. Standard SE fits displayed; phases not searched.}
\end{figure*}

In order to obtain an ED solution on which to base the SE solution of Table
\ref{tab:pole3}, the SP09 solution was used as a starting point, from which a
modified ED result was obtained from the fit to pseudo data, up to a
photon energy of 1.2 GeV. A SE solution was then obtained, starting from ED,
as described above. (Above, and in the following, the notation SE$n$ or ED$n$ 
denotes a fit to $n$ observables.)  We explore these solutions (the ED solution and its
associated SE solutions) in detail below. The $\chi$-squared for the resulting
SE fit (SE4) at 340 MeV is 116 for 144 $d\sigma/d\Omega$, $P$, $\Sigma$,
and $T$ pseudo data. We note here, for subsequent
discussion, that only the $\ell = 0,1$ multipoles depart significantly from a
Born approximation at the $\Delta(1232)$ resonance energy.

An alternative method that can be used at this energy, which reduces
the influence of Watson's theorem, is to first fit the multipoles
for $\pi^+ n$ photoproduction (both real and imaginary parts) for
$\ell=0,1$, using the real Born multipoles for higher waves, to fix the
overall phase. One can then fit the $\pi^0 p$ data separately, again
fitting both the real and imaginary parts of contributing multipoles,
with the $M_{1+}^{\pi^0 p}$ multipole phase fixed, given
the associated $\pi N$ phase (by Watson's theorem, but in this $P_{33}$ partial
wave only) when combined with the previously determined $M_{1+}^{\pi^+
n}$ multipole. This method requires fewer constraints, but results
in multipoles with errors much larger than those given in Table
\ref{tab:pole3}.


\begin{figure*}
\includegraphics[width=0.35\textwidth, keepaspectratio, angle=90]{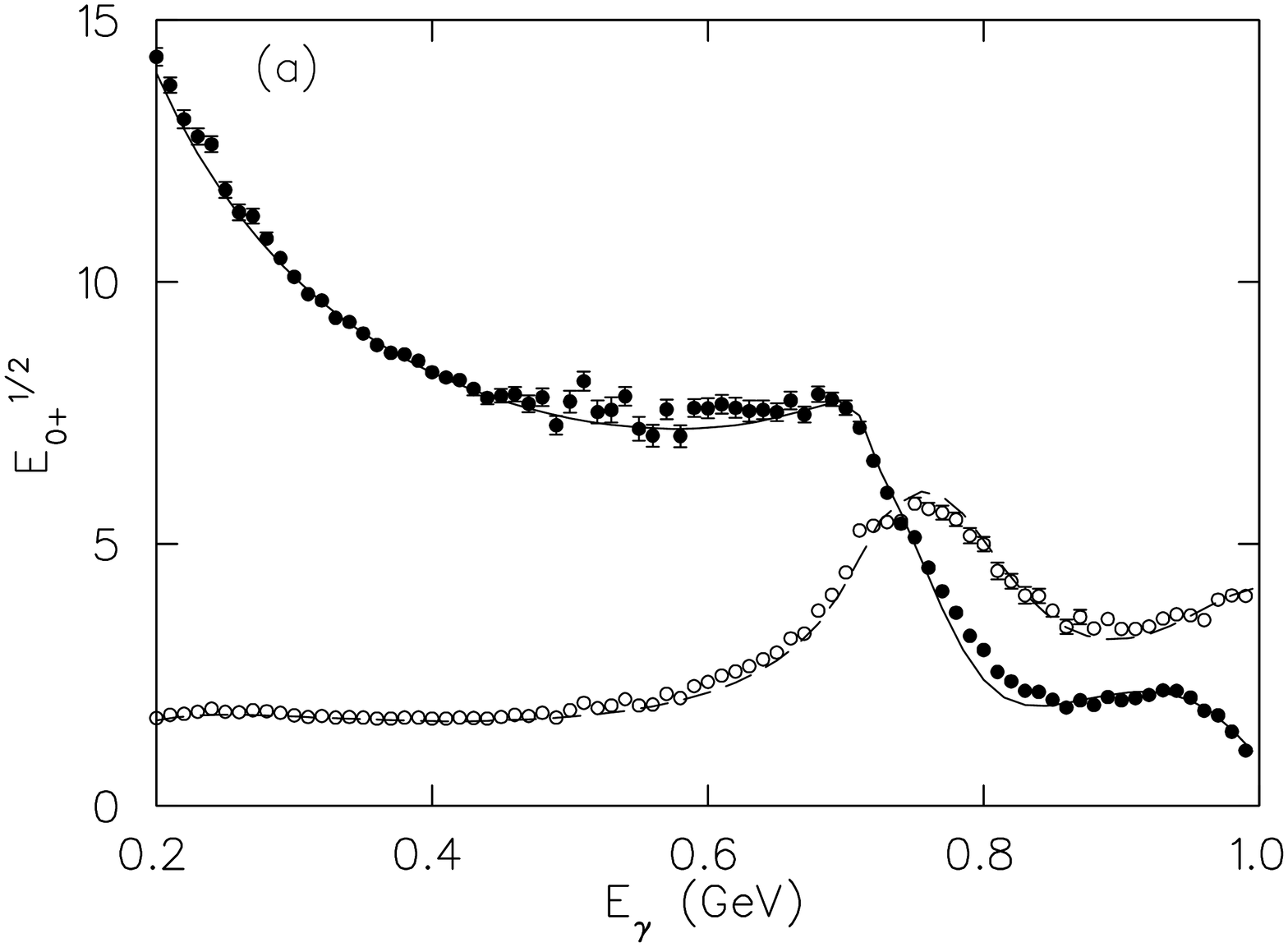}
\hspace*{0.2cm}
\includegraphics[width=0.35\textwidth, keepaspectratio, angle=90]{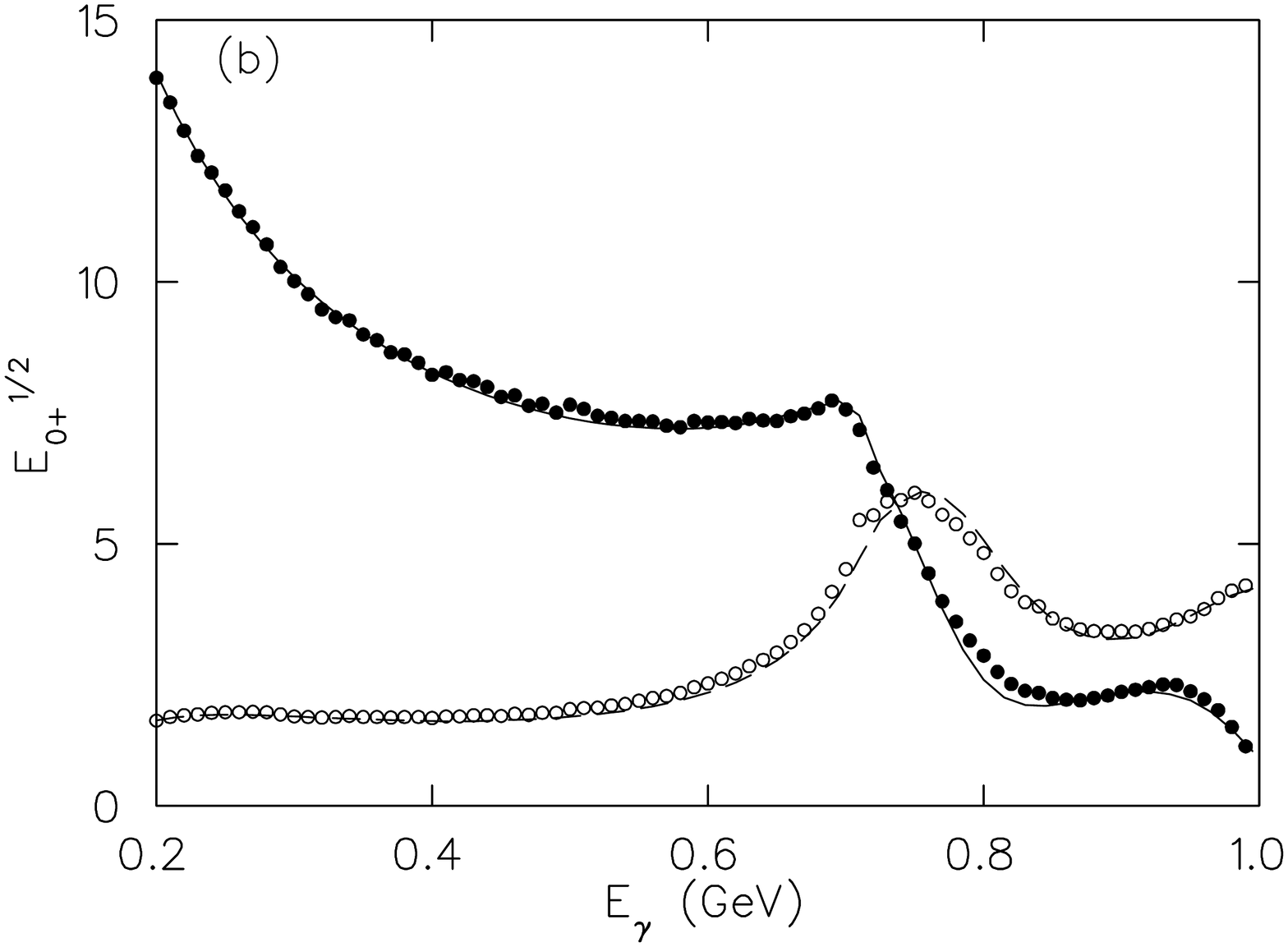}
\caption{\label{fig:pws11} Real and imaginary parts of
the $S_{11}(pE)$ partial wave amplitude ($E_{0+}^{1/2}$).
Notation as in Fig.~\ref{fig:pwp11}.}
\end{figure*}

A form of this two-step method was used in an analysis by Grushin\cite{grushin}
and a recent update\cite{Workman:2010xc} and was found to give multipoles with
phases consistent with Watson's theorem over the photon lab energy range of
$280 < E_\gamma < 420$ MeV. As the $P_{33}$ $\pi N$ partial wave remains
essentially elastic for CM energies $W \lesssim 1450$ MeV ($E_{\gamma} \lesssim
650$ MeV) the method could, in principle, be extended to this energy. This
defines a second energy region, where not all multipoles have phases set by
Watson's theorem, but reliable results can be obtained, with a self-consistent
database of experimental reaction observables. A study by Omelaenko~\cite{omel}
suggests that the number of angular measurements required in order to determine
multipoles up to a given angular momentum is of order $2\ell$ for the
differential cross section, and each single-polarization observable: $P$,
$\Sigma$, and $T$. An additional double polarization measurement, such as $F$
or $G$, is required to fix a remaining discrete ambiguity. Here we are fitting
18 angular pseudo data for each observable, which clearly exceeds the
requirements given in Ref.~\cite{omel}. Note also that the case considered in
Ref.~\cite{omel} was neutral pion photoproduction, and did not depend on the
general use of Watson's theorem nor the charged-pion
channel to determine the overall phase.

In Fig.\eqref{fig:f1}, the SE and ED fits to pseudo data are compared to the
cross section and single-spin asymmetries at 340 MeV. Agreement of the SE and
ED solutions with the pseudo data for $d\sigma/d\Omega$ and the
single-polarization observables, $P$, $\Sigma$, and $T$ is obtained; agreement
of the predicted double-spin asymmetries (not shown) is obtained at a similar
level.  This has been accomplished without fitting the complete set of
experiments required for amplitude reconstruction, as described in
Ref.~\cite{chiang}.

In Fig.\eqref{fig:f2}, this process has been repeated at 600 MeV. The SE fit to
$d\sigma /d\Omega$ and single-spin asymmetries ($P$, $\Sigma$, $T$) at this
energy agree with the pseudo data, with a $\chi$-squared of 89 for 144 data. At
this photon (CM) energy, 600 MeV (1420 MeV), Watson's theorem is not generally
valid in all multipoles. Inelasticity in the $\pi N$ elastic $P_{33}$ partial
wave is, however, small up to 1450 MeV\cite{Arndt:2006bf}. If we assume the
$P_{33}(pM)$ $(M_{1+}^{3/2})$ multipole has the same phase as the $\pi N$
elastic $P_{33}$ amplitude and that the high partial waves for $\pi^+ n$
photoproduction are given by the Born terms, which are real, this should allow
the reconstruction of all relevant multipoles. This was the approach taken in
Ref.\cite{Workman:2010xc}, although at lower energies than 600 MeV. This
expectation is supported by Figs.\eqref{fig:f3} and \eqref{fig:f4} where the
agreement between the displayed curves is of fair quality, though lower than
that of the Fig.\eqref{fig:f2}.

In Fig.\eqref{fig:f3}, for example, the prediction for beam-target observables
from the SE fit is compared to the pseudo data and the ED solution (ED4).
The agreement is generally good but at the level of several standard
deviations with respect to the pseudo data error bars at some angles. These
discrepancies are reflected in the predicted multipoles\cite{SAID-unp}.

The SAID SE procedure
requires that the first ED fit to pseudo data provides the proper phases, as
the subsequent SE search holds these phases constant. 
We can improve the agreement, however, by relaxing the phase constraint on the SE
solution.
In Fig.\eqref{fig:f4}, we
compare phases of the initial solution (SP09) to the ED re-fit (ED4), and the
solution underlying the pseudo data (MD07). The agreement between MD07 and ED4
is generally good up to about 600 MeV, though there are very large differences
between MD07 and SP09 in some multipoles.

In Table \ref{tab:pole6}, the phases of several multipoles are compared at 600
MeV.  Of particular interest are the multipoles $E_{0+}^{1/2}$ and $E_{2-}^{1/2}$,
connected to the $\pi N$
partial waves $S_{11}$ and $D_{13}$, which are known to differ significantly
between MAID and SAID, as can be seen comparing the values from MD07 and SP09.
Note that the ED fit has improved the agreement, but does not match the
phases for the $D_{13}$ multipoles, as shown in Fig.\eqref{fig:f4}, panel (d).
The SE fit to cross section and polarization observables, displayed in
Figs.\eqref{fig:f2} and \eqref{fig:f3}, retains this phase mismatch. In the fit
SE4p, the same set of four observables ($d\sigma/d\Omega$, $P$, $\Sigma$, $T$)
is fitted, while allowing the multipole phases to vary; the result is tabulated
in Table \ref{tab:pole6}. In SE8p, the set of fitted pseudo data has been
expanded to contain beam-target double polarization quantities ($E$, $F$, $G$,
and $H$), again allowing the phase to vary. By allowing phases to vary, we can
now obtain a good reconstruction of the underlying model (MD07) for the
dominant multipoles\cite{SAID-unp}.

The `standard' SE fit results (where the phases are not searched) are
given for two multipoles in Fig.\eqref{fig:pwp11} and
\eqref{fig:pws11}.  The left and right panels display the effect of
adding beam-target observables to the database. While uncertainties
and scatter diminish with additional observables, some slight
systematic deviations are visible. As described above, these can be
reduced by allowing phase variation in the fit.

At much higher energies, where multipoles connected to the $\pi N$ $P_{33}$
partial-wave are no longer constrained in phase, the agreement deteriorates.
However, the phase comparisons in Fig.\eqref{fig:f4} suggest that the ED refit
to pseudo data generated from MD07 obtains the approximate phase of MD07 in
many partial waves. It would be interesting to repeat this comparison with
models other than MAID and SAID to better gauge the model-dependence of fit
results at higher energies.

\section{Conclusion}
\label{sec:conc}

We have performed ED and SE fits to realistic pseudo data generated from the
MD07 ED multipole analysis over a range of photon (CM) energies from 200 (1120)
to 1200 (1770) MeV. As a prelude, we have provided a detailed explanation of
the amplitude definitions and the methods used in SE analysis by the MAID and
SAID groups. This should be useful to those planning experiments and wishing to
compare with the MAID and SAID predictions. It should also clarify the
connections between SE and ED fits, and the data they analyze.

We have conducted a number of fits in order to explore the
model-dependence of SE results in specific energy regions associated
with pion photoproduction. Unlike the amplitude reconstruction
process, which does not yield unique multipole solutions, the method
described here appears to give good results with only fits to
high-quality cross section and single-polarization data for energies
where Watson's theorem is valid. A second region, extending over
energies where the $P_{33}$ $\pi N$ amplitude remains elastic, has
also been found to yield reproducible multipoles. At higher energies,
results become more model-dependent, and further study is required to
determine the reliability of SE multipoles determined via the methods
described above.

\begin{acknowledgments}
This work was supported in part by the U.S.\ Department of Energy
Grant DE-FG02-99ER41110, by the Deutsche Forschungsgemeinschaft (DFG)
via the Sonderforschungsbereich SFB443 and the European
Community-Research Infrastructure Activity under the FP6
(HadronPhysics, RII3-CT-2004-506078) and FP7 (HadronPhysicsII)
programs. LT and MP wish to thank the European Centre for Theoretical
Studies in Nuclear Physics for support during a portion of this
project.
\end{acknowledgments}

\end{document}